\documentclass[aps,pre,twocolumn,superscriptaddress,showpacs]{revtex4}

\usepackage{graphicx}
\usepackage{dcolumn}
\usepackage{bm}
\usepackage{bbm}
\usepackage{psfrag}
\usepackage{multirow,amssymb,amsbsy,amsmath}
\usepackage{stmaryrd}

\newcommand{\I}{\textup{i}}
\newcommand{\E}{\textup{e}}
\newcommand{\D}{\textup{d}}
\newcommand*{\Haver}[1]{\mathopen{\llbracket} #1 \mathclose{\rrbracket}}
\newcommand{\vek}[1]{\bm{#1}}
\newcommand{\dd}{\text{d}}
\newcommand{\dod}[2]{\frac{\dd #1}{\dd #2}}

\newcommand{\pon}[3]{\frac{\partial^{#3}#1}{\partial #2^{#3}}}
\newcommand{\Funktion}[2]{#1\kern-0.2em\left(#2\right)}
\newcommand{\Funktiontxt}[2]{#1(#2)}
\newcommand{\expfkt}[1]{\Funktion{\exp}{#1}}
\newcommand{\expfkttxt}[1]{\Funktiontxt{\exp}{#1}}
\newcommand{\tr}[2][]{\text{Tr}_{#1}\left\{#2\right\}}
\newcommand{\trtxt}[2][]{\text{Tr}_{#1}\{#2\}}
\newcommand*{\bra}[1]{\mathopen{\langle}#1\mathclose{|}}
\newcommand*{\ket}[1]{\mathopen{|}#1\mathclose{\rangle}}
\newcommand*{\sprod}[2]{\mathopen{\langle}#1|#2\mathclose{\rangle}}
\newcommand*{\erwart}[1]{\mathopen{\langle}#1\mathclose{\rangle}}
\newcommand{\refsec}[1]{Sect.~\ref{#1}}
\newcommand{\reffig}[1]{Fig.~\ref{#1}}

\begin{document}

%
%
\title{Application of the Hilbert Space Average Method on Heat Conduction Models}

\author{Mathias Michel}
\affiliation{Institute of Theoretical Physics I, University of Stuttgart, %
             Pfaffenwaldring 57, 70550 Stuttgart, Germany}%
\email{mathias@theo1.physik.uni-stuttgart.de}
\author{Jochen Gemmer}%
\affiliation{Physics Department, University of Osnabr\"uck, %
             Barbarastr.\ 7, 49069 Osnabr\"uck, Germany}%
\author{G\"unter Mahler}
\affiliation{Institute of Theoretical Physics I, University of Stuttgart, %
             Pfaffenwaldring 57, 70550 Stuttgart, Germany}%

\date{\today}

\begin{abstract}
We analyze closed one-dimensional chains of weakly coupled many level systems, by means of the so-called Hilbert space average method (HAM). 
Subject to some concrete conditions on the Hamiltonian of the system, our theory predicts energy diffusion with respect to a coarse-grained description for almost all initial states.
Close to the respective equilibrium we investigate this behavior in terms of heat transport and derive the heat conduction coefficient.
Thus, we are able to show that both heat (energy) diffusive behavior as well as Fourier's law follows from and is compatible with a reversible Schr\"odinger dynamics on the complete level of description. 
\end{abstract}


\maketitle

%
%

%
%
\section{Introduction}
\label{sec:1}

As a central problem of non-equilibrium thermodynamics, heat conduction has been of vital interest since the beginning of the 19th century.
During his extensive investigations on heat transport in solids Fourier discovered around 1807 \cite{Fourier1955} a proportionality between the temperature gradient $\vek{\nabla} T$ and the heat respectively energy current density in the material,
\begin{equation}
  \label{eq:1}
  \vek{j} = - \Lambda \vek{\nabla} T\;,
\end{equation}
where $\Lambda$ is some material constant -- the heat conductivity.
Concentrating on one-dimensional chains of interacting subsystems the temperature gradient can be discretized as $|\vek{\nabla} T|=\Delta T/\Delta L$, where $\Delta T$ is the temperature difference between two adjacent subsystems and $\Delta L$ their distance.
Introducing the current $J=Aj$ through some area $A$, Fourier's law (\ref{eq:1}) can be rewritten as
\begin{equation}
  \label{eq:2}
  J = - \kappa_{\text{th}} \Delta T
\end{equation}
with $\kappa_{\text{th}} = \Lambda A/\Delta L$.
Instead of approaching a global equilibrium state with some temperature $T$, a solid, under some external perturbation by a temperature gradient (introduced by heat baths with different temperatures), enters a local equilibrium state only: the temperature may vary from one macroscopically small but microscopically large part of the system to the other.
The heat transport itself is called \emph{normal} if the conductivity $\kappa_{\text{th}}$ remains finite, whereas a diverging $\kappa_{\text{th}}$ indicates a \emph{ballistic} transport.
Having established a normal diffusive behavior, all further efforts on the topic of heat conduction in the last decades have been directed towards a microscopic foundation of the above phenomenological law.

One of the first concepts addressing the occurrence of regular heat conduction from a microscopic basis is the Peierls-Boltzmann theory \cite{Peierls1955}. 
Essentially, Peierls proposed a modified Boltzmann equation replacing classical particles by quantized quasi particles, the phonons.
He was able to show that in order to obtain normal heat conduction an anharmonicity in the interaction potential was necessary, associated with the well known Umklapp processes of phonon scattering in solids.
Since quantized normal modes are treated here as classical particles, i.e., as being always well localized in configuration as well as momentum space, this approach contains some conceptual shortcomings, though. 

Another microscopic approach to heat conduction is the famous Green-Kubo formula, derived on the basis of linear response theory \cite{Kubo1957,Kubo1991,Mori1956}.
Thereby, the current is taken as the response of the material to an external perturbative potential, originally, the electromagnetic potential.
By a direct mapping of these ideas on thermal transport phenomena (perturbations due to thermal gradients \cite{Luttinger1964}) serious questions remain open, since in this case it is not possible to define a proper potential in the Hamiltonian of the system resulting in a temperature gradient.

To account for the problem of defining a proper potential for a thermal perturbation, there have been efforts to model the system coupled to several heat baths in terms of an open quantum system \cite{Saito2002,Michel2003,Michel2005}.
These methods reveal a normal transport behavior even in very small quantum systems, but are numerically challenging, especially for larger systems, since the investigations are done in the Liouville space instead of the Hilbert space.

All of the above microscopic approaches to heat conduction assume that the system under consideration is perturbed by some external environment (e.g.\ a scenario, where a solid is heated at one side cooled at the other).
As already predicted by Einstein in his considerations on Brownian motion \cite{Einstein1905,Einstein1908}, the diffusion constant appearing in a local equilibrium steady state scenario of heat conduction should equal the decay constant for the relaxation from a non-equilibrium state to the global equilibrium in the same system.
In the latter case there is no further external perturbation involved.
In the same spirit we will introduce yet another approach to heat transport within quantum systems.

Recently, the relaxation behavior of small quantum systems has been considered within the theory of quantum thermodynamics \cite{GemmerOtte2001,BorowskiGemmer2003,Henrich2005,Gemmer2004}, calling for a direct solution of the Schr\"odinger equation.
Based on the Hilbert space Average Method (HAM) \cite{Gemmer2004,Gemmer2003,Gemmer2005I,Gemmer2005II}, the complicated Schr\"odinger dynamics of a multi level system may be reduced to a set of simple rate equations.
In the present paper we will discuss the application of HAM on heat conducting model systems.
Furthermore, we will discuss the connection of heat diffusive behavior with energy diffusion and present a comparison of the results of HAM with the full numerical solution of the Schr\"odinger equation.
The goal of the present investigation is to show how heat respectively energy diffusive behavior may emerge from first principles -- here the Schr\"odinger equation \cite{Michel2005II}.

%
%
\section{Single Band Model}
\label{sec:2}

\begin{figure}
  \centering 
  \includegraphics[width=7cm]{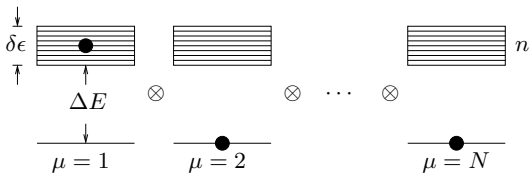}
  \caption{Heat conduction model: $N$ coupled subunits with ground level and band of $n$ equally distributed levels ($\Delta E =1$). Black dots indicate the initial state used.}
\label{fig:1}  
\end{figure}

The class of systems we are going to analyze first is depicted in Fig.~\ref{fig:1}, and described by the Hamiltonian  
\begin{equation}
  \label{eq:3}
  \hat{H} = \sum_{\mu=1}^{N} \hat{H}_{\text{loc}}(\mu) 
          + \lambda\sum_{\mu=1}^{N-1} \hat{V}(\mu,\mu+1) \;.
\end{equation}
Here $N$ identical subunits according to the same local Hamiltonian $\hat{H}_{\text{loc}}(\mu)$ are assumed to have a non-degenerate ground state and a band of $n$ exited states each, equally distributed over some band width $\delta\epsilon$ such that the band width is small compared to the local energy gap $\Delta E$ (see \reffig{fig:1}, $\delta\epsilon\ll\Delta E$, $\delta\epsilon$ in units of $\Delta E$).
These subunits are coupled by an energy exchanging next neighbor interaction $\hat{V}(\mu,\mu+1)$, chosen to be a (normalized) random hermitian matrix allowing for any possible transition such as to avoid any bias. 
Our results will turn out to be independent of the exact form of this matrix.
We choose the next neighbor coupling to be weak compared to the local gap ($\lambda\ll\Delta E$, $\lambda$ in units of $\Delta E$).
This way the full energy is approximately given by the sum of the local energies and those are approximately given by $\erwart{\hat{H}_{\text{loc}}(\mu)} \approx \Delta E P_{\mu}$, where $P_{\mu}$ is the probability to find the $\mu$-th subsystem in its excited state.
One of the big advantages of the present model is the clear partition of the Hamiltonian in a local and an interaction part, respectively.
This partition allows for a unique definition of both a local energy or even a local temperature as well as a proper current.
However, note that the presented method itself does not need such a special kind of model system but is in principle also applicable to more general situations. 

We now restrict ourselves to the single excitation subspace, i.e., all states with one subsystem excited (irrespective of which state within the band), all others in the ground state.
Let us introduce a set of projection operators $\hat{P}_{\mu,\mu}$ in this subspace, projecting out the set of states corresponding to the $\mu$-th subsystem occupying its excitation band, rather than its ground state (these operators feature the unit operator of dimension $n$ for the excited band of the $\mu$-th system else zeros).
Thus, the total system state is given by 
\begin{equation}
  \label{eq:4}
  \ket{\psi}
  = \sum_{\mu} \hat{P}_{\mu,\mu}\ket{\psi}
  = \sum_{\mu} \ket{\psi_{\mu}}\;,
\end{equation}
where none of the $\ket{\psi_{\mu}}$ is normalized individually.
Furthermore, one may introduce also some standard transition operators $\hat{P}_{\mu,\nu}$ (unit operator at position $(\mu,\nu)$).
According to the properties of transition operators we find $\hat{P}_{\mu,\nu}\hat{P}_{\mu',\nu'}=\delta_{\nu\mu'}\hat{P}_{\mu,\nu'}$ and hence
\begin{equation}
  \label{eq:5}
  \hat{P}_{\mu+1,\mu}\hat{P}_{\mu,\mu}=\hat{P}_{\mu+1,\mu+1}\hat{P}_{\mu+1,\mu}
  \;.
\end{equation}
Both, the local and the interaction part of the Hamiltonian (\ref{eq:3}) in our reduced Hamiltonian model can now be written in terms of these new operators as
\begin{align}
  \label{eq:6}
  \hat{H}_{\text{loc}} (\mu)
  &= h_{\mu,\mu}\hat{P}_{\mu,\mu}\;,\\
  \label{eq:7}
  \hat{V}(\mu,\mu+1)
  &= \big(
     v_{\mu,\mu+1}\hat{P}_{\mu,\mu+1}+v_{\mu+1,\mu}\hat{P}_{\mu+1,\mu}\big)\;,
\end{align}
with a local diagonal Hamiltonian matrix $h_{\mu,\mu}$ and a hermitian interaction matrix $v_{\mu,\mu+1}=v_{\mu+1,\mu}^{\dagger}$.

%
%
\section{Time Evolution}
\label{sec:3}

The dynamics of the model system defined by the Hamiltonian (\ref{eq:3}) respectively (\ref{eq:6}) and (\ref{eq:7}) is governed by the Schr\"odinger equation.
Here we will mainly be interested in a coarse-grained description suppressing the information about the state index within the respective bands, concentrating, e.g., on the total probability of occupying a band or not, only.

After transforming the Hamiltonian to the interaction picture we perform a second order Dyson expansion of the time evolution for the short time step $\tau$
\begin{equation}
  \label{eq:8}
  \ket{\psi(\tau)}
  \approx\big(\hat{1}-\frac{\I}{\hbar}\hat{U}^{(1)}
                     -\frac{1}{\hbar^2}\hat{U}^{(2)}\big)
  \ket{\psi(0)}
  = \hat{D}\;\ket{\psi(0)}\;,
\end{equation}
with the two time evolution operators 
\begin{align}
  \label{eq:9}
  \hat{U}^{(1)}(\tau) &= \int_0^{\tau}\D\tau'\,\hat{V}(\tau')\;, 
  \\
  \label{eq:10}
  \hat{U}^{(2)}(\tau) &= \int_0^{\tau}\D\tau'\int_0^{\tau'}\D\tau''
                    \,\hat{V}(\tau') \,\hat{V}(\tau'')\;.
\end{align}

A pertinent quantity in the context of heat conduction will be the change of local energy in one subunit $\mu$ in the time interval $\tau$ given by the probability 
\begin{equation}
  \label{eq:11}
  P_{\mu}(\tau)
  =\sprod{\psi_{\mu}(\tau)}{\psi_{\mu}(\tau)}
  =\bra{\psi(\tau)}\hat{P}_{\mu,\mu}\hat{P}_{\mu,\mu}\ket{\psi(\tau)}\;.
\end{equation}
Using the short time evolution operator (\ref{eq:8}), one gets 
\begin{equation}
  \label{eq:12}
  P_{\mu}(\tau)
  =\bra{\psi(0)}\hat{D}^{\dagger}\hat{P}_{\mu,\mu}\hat{P}_{\mu,\mu}\hat{D}\ket{\psi(0)}
  :=\bra{\psi(0)}\hat{A}_{\mu}\ket{\psi(0)}\;.
\end{equation}
For simplicity we suppress in the following the explicit notation of the time dependence of the operators, states and probabilities.
According to the definition of $\hat{D}$ we find in second order
\begin{align}
  \label{eq:13}
  \hat{A}_{\mu}
   =& \hat{P}_{\mu,\mu}\hat{P}_{\mu,\mu}
   +  \frac{\I}{\hbar} \hat{U}^{(1)}\hat{P}_{\mu,\mu}\hat{P}_{\mu,\mu}
   -  \frac{\I}{\hbar} \hat{P}_{\mu,\mu}\hat{P}_{\mu,\mu}\hat{U}^{(1)}
   \notag\\
   &- \frac{1}{\hbar^2}\hat{P}_{\mu,\mu}\hat{P}_{\mu,\mu}\hat{U}^{(2)}
    - \frac{1}{\hbar^2}(\hat{U}^{(2)})^{\dagger}\hat{P}_{\mu,\mu}\hat{P}_{\mu,\mu}
   \notag\\
   &+ \frac{1}{\hbar^2}\hat{U}^{(1)}\hat{P}_{\mu,\mu}\hat{P}_{\mu,\mu}\hat{U}^{(1)}\;.
\end{align}
(Note that $\hat{U}^{(1)}$ is hermitian due to the hermiticity of the interaction).

Due to the additive representation of the interaction (\ref{eq:7}) and the definition of the first oder time evolution operator (\ref{eq:9}) one may find the analogous off-diagonal block form for the first oder time-evolution operator
\begin{equation}
  \label{eq:14}
  \hat{U}^{(1)} 
  =  \sum_{\mu=1}^{N-1} \big(
     u_{\mu,\mu+1}^{(1)}\hat{P}_{\mu,\mu+1}
  +  u_{\mu+1,\mu}^{(1)}\hat{P}_{\mu+1,\mu}\big)\;.
\end{equation}
The time dependent coefficients $u$ contain the integration of the interaction matrix $v$ over $\tau'$ (cf.\ (\ref{eq:7})).

Unfortunately, the second oder time evolution operator (\ref{eq:10}) has a more complicated structure in terms of the transition operators.
Introducing the interaction (\ref{eq:7}) into (\ref{eq:10}) (using the property (\ref{eq:5})) one finds
\begin{align}
  \label{eq:15}
  \hat{U}^{(2)} =& \int_0^{\tau}\D\tau'\int_0^{\tau'}\D\tau''
    \sum_{\mu}
    \big(
    v_{\mu,\mu+1}(\tau')v_{\mu+1,\mu}(\tau'') \hat{P}_{\mu,\mu}
    \notag\\
    &+ v_{\mu+1,\mu}(\tau')v_{\mu,\mu+1}(\tau'') \hat{P}_{\mu+1,\mu+1}
    \notag\\
    &+ v_{\mu,\mu+1}(\tau')v_{\mu+1,\mu+2}(\tau'')\hat{P}_{\mu,\mu+2}
    \notag\\
    &+ v_{\mu+1,\mu}(\tau')v_{\mu,\mu-1}(\tau'')\hat{P}_{\mu+1,\mu-1}\big)\;.
\end{align}
This operator decomposes into a diagonal and an off-diagonal part $\hat{U}^{(2)}=\hat{U}^{(2)}_{\text{diag}} + \hat{U}^{(2)}_{\text{off}}$ with
\begin{align}
  \label{eq:16}
  \hat{U}^{(2)}_{\text{off}}
  &= \sum_{\mu}\big(u^{(2)}_{\mu-1,\mu+1}\hat{P}_{\mu-1,\mu+1}
                  + u^{(2)}_{\mu+1,\mu-1}\hat{P}_{\mu+1,\mu-1}\big)\;,
  \notag\\
  \hat{U}^{(2)}_{\text{diag}}
  &= \sum_{\mu} u^{(2)}_{\mu,\mu} \hat{P}_{\mu,\mu}\;,
\end{align}
and a suitable definition of the coefficients $u^{(2)}$.
By using (\ref{eq:5}) one may convince oneself of the operator identities
\begin{align}
  \label{eq:17}
  \hat{U}^{(1)}\hat{P}_{\mu,\mu}
  &= (\hat{P}_{\mu-1,\mu-1}+\hat{P}_{\mu+1,\mu+1})\hat{U}^{(1)}\;,
  \\
  \label{eq:18}
  \hat{P}_{\mu,\mu} \hat{U}^{(2)}_{\text{off}} 
  &= \hat{U}^{(2)}_{\text{off}} 
    \big(\hat{P}_{\mu+2,\mu+2}+\hat{P}_{\mu-2,\mu-2} \big)\;,
  \\
  \label{eq:19}
  \hat{P}_{\mu,\mu} \hat{U}^{(2)}_{\text{diag}}
  &= \hat{U}^{(2)}_{\text{diag}} \hat{P}_{\mu,\mu}\;.
\end{align}
By means of the expansion of the time-evolution operators in terms of projectors and transition operators and those operator identities one is able to reformulate (\ref{eq:13}) by exchanging time evolution operators and projectors as
\begin{align}
  \label{eq:20}
  \hat{A}_{\mu}
  =& \phantom{+}\hat{P}_{\mu,\mu}\hat{P}_{\mu,\mu}
  \notag\\
  &+ \frac{\I}{\hbar} 
     (\hat{P}_{\mu-1,\mu-1}+\hat{P}_{\mu+1,\mu+1})\hat{U}^{(1)}
     \hat{P}_{\mu,\mu}
  \notag\\
  &- \frac{\I}{\hbar}
     \hat{P}_{\mu,\mu}
     \hat{U}^{(1)}(\hat{P}_{\mu-1,\mu-1}+\hat{P}_{\mu+1,\mu+1})
  \notag\\
  &+ \frac{\I}{\hbar}
     \hat{P}_{\mu-1,\mu-1}
     \big(\hat{U}^{(1)}\big)^2
     (\hat{P}_{\mu-1,\mu-1}+\hat{P}_{\mu+1,\mu+1})
  \notag\\
  &+ \frac{\I}{\hbar}
     \hat{P}_{\mu+1,\mu+1}
     \big(\hat{U}^{(1)}\big)^2
     (\hat{P}_{\mu-1,\mu-1}+\hat{P}_{\mu+1,\mu+1})
  \notag\\
  &- \frac{\I}{\hbar}
     \hat{P}_{\mu,\mu}
     \Big(\hat{U}^{(2)}_{\text{diag}}
         +\big(\hat{U}^{(2)}_{\text{diag}}\big)^{\dagger}\Big)
     \hat{P}_{\mu,\mu}
  \notag\\
  &- \frac{\I}{\hbar}
     \hat{P}_{\mu,\mu}
     \hat{U}^{(2)}_{\text{off}}
     (\hat{P}_{\mu-2,\mu-2}+\hat{P}_{\mu+2,\mu+2})
  \notag\\
  &- \frac{\I}{\hbar}
     (\hat{P}_{\mu-2,\mu-2}+\hat{P}_{\mu+2,\mu+2})
     \big(\hat{U}^{(2)}_{\text{off}}\big)^{\dagger}
     \hat{P}_{\mu,\mu}
\end{align}
Since $P_{\mu}=\bra{\psi}\hat{A}_{\mu}\ket{\psi}$ is a probability we get, furthermore, the normalization condition $\sum_{\mu} \bra{\psi}\hat{A}_{\mu}\ket{\psi}=1$.
This condition is already fulfilled in the zeroth order of all probabilities $\sum_{\mu} \bra{\psi}\hat{P}_{\mu,\mu}^2\ket{\psi}=1$. 
Due to this fact the sum over all other orders has to vanish.
All first order terms together cancel out each other directly.
Remaining second oder terms lead to the additional conditions 
\begin{align}
  \label{eq:21}
  \bra{\psi_{\mu-1}}\hat{U}^{(2)}_{\text{off}}
         +\big(\hat{U}^{(2)}_{\text{off}}\big)^{\dagger}\ket{\psi_{\mu+1}}
  &= \bra{\psi_{\mu-1}}\big(\hat{U}^{(1)}\big)^2 \ket{\psi_{\mu+1}}\;,
  \notag\\
  \bra{\psi_{\mu}}\hat{U}^{(2)}_{\text{diag}}
         +\big(\hat{U}^{(2)}_{\text{diag}}\big)^{\dagger}\ket{\psi_{\mu}}
  &= 2 \bra{\psi_{\mu}}\big(\hat{U}^{(1)}\big)^2 \ket{\psi_{\mu}}\;.
\end{align}

%
%
\section{Energy Transport in the Single Band Model}
\label{sec:4}

According to (\ref{eq:11}) the expectation value of the operator (\ref{eq:20}) defines the time dependence of the probability $P_{\mu}$.
We approximate this expectation value by an average over some appropriate Hilbert space compartment -- say the mean of the expectation value over all reachable states -- called the Hilbert space average $\Haver{\cdots}$.

The Hilbert space average method (HAM) is, in es\-sence, a technique to guess the value of some quantity defined as a function of the full system state $\ket{\psi}$ if $\ket{\psi}$ itself is not known in full detail.
Here it is used to produce a guess for the expectation values $\bra{\psi}\hat{A}_{\mu}\ket{\psi}$ if the only information about $\ket{\psi}$ is the given set of initial expectation values $\bra{\psi}\hat{P}_{\mu,\mu}^2\ket{\psi} = P_{\mu}(0)$. 
Such a statement naturally has to be a guess, since there are, in general, many different $\ket{\psi}$ that are in accord with the given set of $P_{\mu}(0)$ but possibly produce different values for $\bra{\psi}\hat{A}_{\mu}\ket{\psi}$. 
Thus, the key question for the reliability of this guess is whether the distribution of the $\bra{\psi}\hat{A}_{\mu}\ket{\psi}$ produced by the respective set of $\ket{\psi}$'s is broad or whether almost all those $\bra{\psi}\hat{A}_{\mu}\ket{\psi}$ are approximately equal. 
It can be shown that if the spectral width of $\hat{A}_{\mu}$ is not too large and  $\hat{A}_{\mu}$ is high-dimensional almost any $\ket{\psi}$ yields an expectation value close to the mean of the distribution of the $\bra{\psi}\hat{A}_{\mu}\ket{\psi}$ and thus the HAM guess will be reliable \cite{Gemmer2004}. 
To find that mean one has to average with respect to the  $\ket{\psi}$'s. 
This is called a Hilbert space average and denoted as 
\begin{equation}
  \label{eq:22}
  P_{\mu}(\tau)
  = \bra{\psi(0)}\hat{A}_{\mu}\ket{\psi(0)}
  \approx \Haver{\bra{\psi}\hat{A}_{\mu}\ket{\psi}}\;.
\end{equation}
Such averages are well known and derived in all detail in \cite{Gemmer2004,Gemmer2005I}. For any hermitian operator $\hat{A}$ one finds 
\begin{equation}
  \label{eq:23}
  \Haver{\bra{\psi_{\mu}}\hat{A}\ket{\psi_{\nu}}}
  = 
  \begin{cases}
    \frac{\sprod{\psi_{\mu}}{\psi_{\mu}}}{d_{\mu}}\;\trtxt[\mu]{\hat{A}} & \mu=\nu\\
    0 & \text{else}\;,
  \end{cases}
\end{equation}
with the partial traces in the respective subspace $\mu$ with dimension $d_{\mu}$ (note $\sprod{\psi_{\mu}}{\psi_{\mu}}\neq 1$).

Plugging in the operator $\hat{A}_{\mu}$ from (\ref{eq:20}) into the probability (\ref{eq:12}), approximating the expectation values by the Hilbert space average and furthermore using the operator identities (\ref{eq:21}), one finds for the probability 
\begin{align}
  \label{eq:24}
  P_{\mu} (\tau)
  = \sprod{\psi_{\mu}}{\psi_{\mu}}
  &-\frac{2}{\hbar^2}
     \Haver{\bra{\psi_{\mu}}\big(\hat{U}^{(1)}\big)^2\ket{\psi_{\mu}}}
  \notag\\
  &+ \frac{1}{\hbar^2}
     \Haver{\bra{\psi_{\mu+1}}\big(\hat{U}^{(1)}\big)^2\ket{\psi_{\mu+1}}}
  \notag\\
  &+\frac{1}{\hbar^2}
     \Haver{\bra{\psi_{\mu-1}}\big(\hat{U}^{(1)}\big)^2\ket{\psi_{\mu-1}}}\;.
\end{align}
Furthermore, using (\ref{eq:23}) with the dimension $d_{\mu}=n$ in each subspace, we end up with the probability
\begin{align}
  \label{eq:25}
  P_{\mu}(\tau) 
  =& \sprod{\psi_{\mu}}{\psi_{\mu}} 
   -\frac{2}{\hbar^2} \frac{\sprod{\psi_{\mu}}{\psi_{\mu}}}{n}\;
                       \tr[\mu]{\big(\hat{U}^{(1)}\big)^2}
  \notag\\
   &+\frac{1}{\hbar^2} \frac{\sprod{\psi_{\mu+1}}{\psi_{\mu+1}}}{n}\;
                       \tr[\mu+1]{\big(\hat{U}^{(1)}\big)^2}
  \notag\\
   &+\frac{1}{\hbar^2} \frac{\sprod{\psi_{\mu-1}}{\psi_{\mu-1}}}{n}\;
                       \tr[\mu-1]{\big(\hat{U}^{(1)}\big)^2}\;.
\end{align}

A first order approximation of these traces is straightforward and done like in the derivation of Fermi's Golden Rule (cf.~\cite{Gemmer2004,Gemmer2005I}) getting here
\begin{equation}
  \label{eq:26}
  \tr[\mu]{\big(\hat{U}^{(1)}\big)^2} 
  \approx \frac{2\pi\hbar\lambda^2n^2\tau}{\delta\epsilon}\;.
\end{equation}
Defining the decay constant 
\begin{equation}
  \label{eq:27}
  \kappa := \frac{2\pi\lambda^2 n}{\hbar\delta\epsilon}
\end{equation}
and remembering that $\sprod{\psi_{\mu}}{\psi_{\mu}}=P_{\mu}(0)=P_{\mu}$ one eventually finds
\begin{equation}
  \label{eq:28}
  P_{\mu}(\tau) 
  = P_{\mu} -  \kappa\Big( 2 P_{\mu}  
  - P_{\mu+1} - P_{\mu-1} \Big)\tau\;.
\end{equation}

The above approximation for the trace of the square of the time-evolution operator (\ref{eq:26}) is only valid within a small time interval like for Fermi's Golden Rule implying the two conditions (see \cite{Gemmer2005I})
\begin{equation}
  \label{eq:29}
  \frac{\lambda}{\delta\epsilon}n\geq\frac{1}{2}\;,
  \quad
  \frac{\lambda^2}{\delta\epsilon^2}n\ll 1\;.
\end{equation}
As long as these conditions are fulfilled the given second order time evolution approximation is valid and one may iterate (\ref{eq:28}) in the limit of $\tau$ being extremely small to find the \emph{rate equation} for the system. 
Since the system is a finite chain of $N$ subsystems we get a slightly different equation for the first and the last system in the chain.
The complete system of differential equations, thus, reads $(2<\mu<N-1)$
\begin{align}
  \label{eq:30}
  \dod{P_{1}}{t} 
  &= -\kappa \Big( P_{1} - P_{2} \Big)\;,
  \\
  \label{eq:31}
  \dod{P_{\mu}}{t} 
  &= -\kappa \Big( 2 P_{\mu} - P_{\mu+1} - P_{\mu-1} \Big)\;,
  \\
  \label{eq:32}
  \dod{P_{N}}{t} 
  &= -\kappa \Big( P_{N} - P_{N-1} \Big)\;.
\end{align}
Under the conditions (\ref{eq:29}) the system should follow this rate equation thus establishing a statistical behavior from pure Schr\"odingerian dynamics!

From (\ref{eq:31}) one finds that the change of the probability in the $\mu$th subsystem only depends on its left and right neighbor.
Therefore, if the above criteria (\ref{eq:29}) are fulfilled, so that the system indeed follows the respective rate equation, we may concentrate on the minimal model of two subunits only, without loss of generality. 
Because of the special next neighbor relationship of the above dynamical equation the results of the minimal model can be generalized to the behavior of a chain of a large number of such subsystems.  

%
%
\section{Numerical Results}
\label{sec:4a}

\begin{figure}
  \centering 
  \includegraphics[width=8cm]{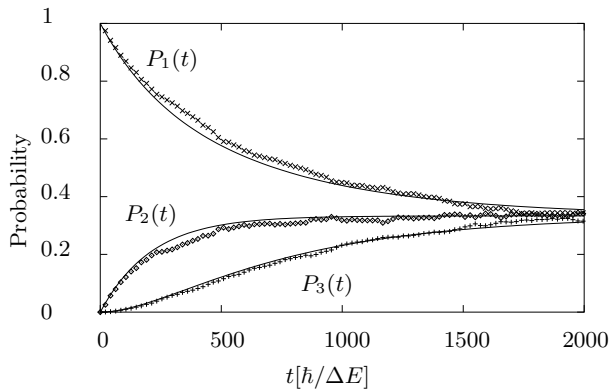}
  \caption{Probability to find the excitation in the $\mu=1,2,3$ system. Comparison of the HAM prediction (lines) and the exact Schr\"odinger solution (dots). ($N=3$, $n=500$, $\lambda=5\cdot 10^{-5}$, $\delta\epsilon=0.05$)} 
\label{fig:2}
\end{figure}

\begin{figure}
  \centering 
  \includegraphics[width=8cm]{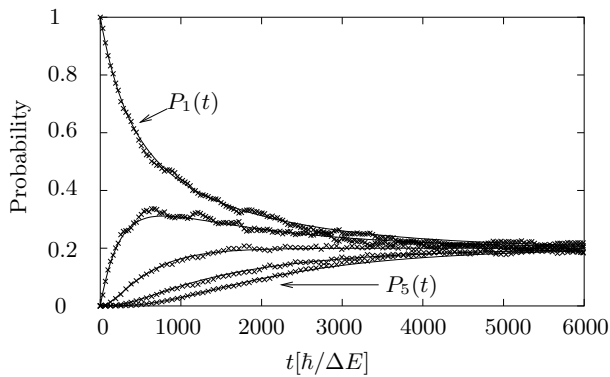}
  \caption{Probability to find the excitation in the $\mu=1,2,3,4,5$ system. Comparison of the HAM prediction (lines) and the exact Schr\"odinger solution (dots). ($N=5$, $n=500$, $\lambda=5\cdot 10^{-5}$, $\delta\epsilon=0.05$)} 
\label{fig:2a}
\end{figure}

That such a behavior as described in the last Sections is, indeed, feasible can be demonstrated by a comparison of the solution of the above rate equation with the exact solution of the full system Schr\"odinger equation. 
For this reason we consider a model consisting of three subunits with $n=500$ levels each in the exited band of width $\delta\epsilon=0.05$ coupled by a random energy exchanging next neighbor coupling $\lambda=5\cdot 10^{-5}$.
For these parameters the two criteria (\ref{eq:29}) are fulfilled.
Initially we start with an excitation in the first system, all other systems being in their ground state, observing the redistribution of energy.
The comparison for $N=3$ in \reffig{fig:2} shows a very good accordance between the solutions of these two different dynamical descriptions -- both the solution of the full Schr\"odinger equation as well as the rate equation (\ref{eq:30})-(\ref{eq:32}).
Furthermore, \reffig{fig:2a} shows the same comparison for a chain of increased length $N=5$.
Again, one finds a very good accordance.

\begin{figure}
  \centering 
  \includegraphics[width=8cm]{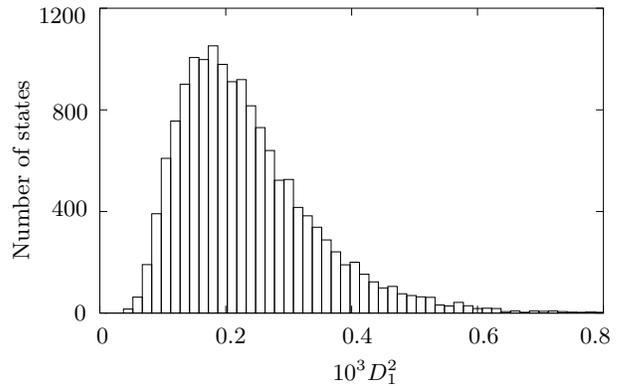}
  \caption{Deviation of HAM from the exact solution ($D^2_1\leq 1$ least squares) for 15000 equally distributed initial states ($N=2$, $n=500$).}
\label{fig:3}  
\end{figure}
To investigate the accuracy of the HAM for, e.g., the probability in the first system $P_1(t)$ we introduce $D^2_1$, as the time-averaged quadratic deviation of the exact (Schr\"o\-din\-ger) result for $P_1(t)$ from the HAM prediction
\begin{equation}  
  \label{eq:33}
  D^2_1=\frac{1}{5\tau}
  \int_0^{5\tau}(P_1^{\text{HAM}}(t)-P_1^{\text{exact}}(t))^2\D t\;.
\end{equation}
To check the validity of HAM with respect to different pure initial states we have computed $D^2_1$ for 15000 equally distributed initial states, all corresponding to a two subunit system with one subunit in the upper band and the other subunit in its ground state.
The results are condensed into the histogram shown in Fig.~\ref{fig:3}.
Obviously, almost all initial states lead to approximately the same small deviation. 
Thus a good agreement of HAM with the exact result may safely be expected regardless of HAM being a best guess only.
(In fact this is the essence of thermodynamical statements: They hold typically, but not always.) 

Furthermore, to analyze how big this ``typical deviation'' is, we have computed $D^2_1$ for a $N=3$ system with the first subunit initially in an arbitrary excited state, for different numbers of states $n$ in the bands. 
\begin{figure}
  \centering 
  \includegraphics[width=8cm]{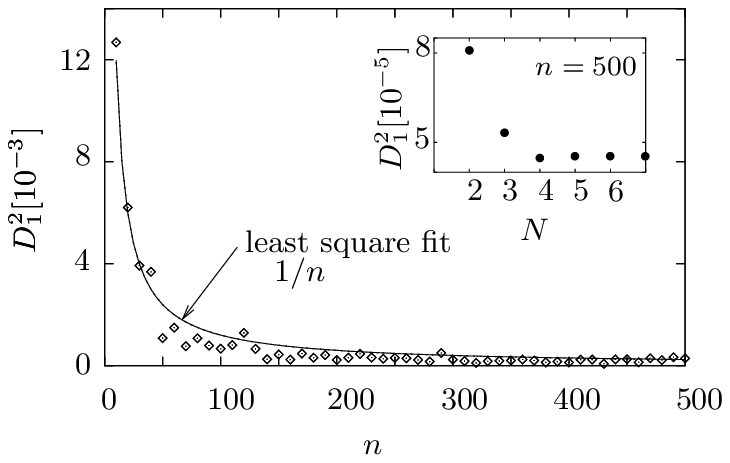}
  \caption{Deviation of HAM from the exact solution ($D^2_1$ least squares): dependence of $D^2_1$ on $n$ for $N=3$. Inset: dependence of $D^2_1$ on $N$ for $n=500$.}
\label{fig:3a}  
\end{figure}
As shown in  Fig.~\ref{fig:3a}, the deviation scales like $1/n$ with the band size, i.e., vanishes in the limit of high dimensional subunits. 
This behavior does not come as a complete surprise since it has theoretically been conjectured and numerically verified in the context of equilibrium fluctuations of non-Markovian systems \cite{Gemmer2004,BorowskiGemmer2003}.
The inset of Fig.~\ref{fig:3a} shows that $D_1^2$ goes down also with increasing number of subunits $N$ but then levels off. 
Altogether the new method HAM appears to be applicable even down to moderately sized subsystems, i.e., not too few levels $n$ in the upper band.
Thus, as follows from \reffig{fig:3}, a system with only a single excited level would lead to large deviations from the rate equation behavior.  

Using instead of pure initial states mixed states (which are typical in the context of thermodynamical phenomena) a drastic further reduction of $D^2$ can be expected.
This is due to the fact that pure state fluctuations may cancel partially if added together.

%
%
\section{Energy Diffusion Constant}
\label{sec:5}

We concentrate in the following on a model (cf.~\reffig{fig:1}) for which the criteria (\ref{eq:29}) are fulfilled.
In essence, the used subunits should be weakly coupled to fulfill the given criteria, i.e., the interaction between the subunits has to be small compared to the local energy.
The dynamical behavior is then approximately governed by the rate equation (\ref{eq:30})-(\ref{eq:32}).
As argued in the last paragraph of \refsec{sec:4}, it is enough to consider a minimal model of $N=2$ subunits, without loss of generality.

This minimal system behaves according to the rate equation (\ref{eq:30}) with the decay constant (\ref{eq:27}).
The respective equation for the second system is simply obtained by an exchange of the indices in (\ref{eq:30}).

In order to compute the energy diffusion constant in the given model system, we have to consider the energy current inside the system.
The energy current is defined by the change of the internal energy $U_{\mu}$ in the two subunits
\begin{equation}
  \label{eq:34}
  J_{1,2} 
  = \frac{1}{2}\left(\dod{U_{1}}{t}-\dod{U_{2}}{t}\right)\;.
\end{equation}
The total internal energy in such a subunit is given by the probability to be in the excited band of the subunit times the width of the energy gap, $U_{\mu}=\Delta E P_{\mu}$.
Thus, we may reformulate the current due to the change of the probability being in the excited band as
\begin{equation}
  \label{eq:35}
  J_{1,2} 
  = \frac{\Delta E}{2}\left(\dod{P_{1}}{t}-\dod{P_{2}}{t}\right)\;.
\end{equation}
The change of the probabilities in time are given by the rate equation (\ref{eq:30}) so that
\begin{equation}
  \label{eq:36}
  J_{1,2} 
  = -\kappa\Delta E\big(P_2-P_1\big)\;.
\end{equation}
For a larger system with $N\geq 2$ the consideration is quite similar and the result in principle the same.
Of course, the current definition is a little bit more complicated, since the energy relaxation in a central system leads to a current to both sides.
But by properly splitting up the result into a current to the left respectively right system leads to the same dependence of the current between subunit $\mu$ and $\mu+1$ on the respective probability difference. 

According to (\ref{eq:36}) the current is a linear function of the probability difference (gradient) respectively \emph{energy difference (gradient)} inside the system.
This refers to a statistical behavior since energy diffuses through the system  only due to the energy distribution in the system.
Therefore we may extract the energy diffusion constant from the above equation, identifying the parameter $\kappa$ with the transport coefficient given in (\ref{eq:27}).
This energy diffusion constant only depends on the coupling strength of the subunits $\lambda$ and on the state density of the excited band $n/\delta\epsilon$.
(Note that in the model at hand the energy gradient is discretized as an energy difference, because of the discrete structure of the model, as already stated in \refsec{sec:1} [cf.~(\ref{eq:2})].)

%
%
\section{Heat Transport in a Multi Band Model}
\label{sec:6}

Up to now we have restricted ourselves to the energy transport within the system on the route from a non-equilibrium initial state to a global equilibrium final state.
Of course this could be translated into heat transport insofar as heat is nothing but thermal energy.
However, the states far away from equilibrium preclude a straight-forward interpretation:
In the example of \reffig{fig:2} we have prepared the system initially in a state with the first system completely in its exited band, all others being in their ground states and, indeed, the system behaves statistically.
By a simple translation to temperatures by fitting to a Boltzmann distribution this initial state would show a really strange temperature profile:
one finds the first system at negative temperature (completely excited, i.e., inversion and therefore negative, infinite temperature), all others at zero temperature.
Nobody would expect a normal heat conduction behavior in such a far from equilibrium situation.
Nevertheless, even in this case we have confirmed a statistical relaxation of energy (see \reffig{fig:2}).

To investigate a situation more appropriate for heat transport we should use an initial state already near the final global equilibrium state, with only a small temperature gradient.
We first start with a pure state observing the decay of such a state by solving the Schr\"odinger equation only.
Finally, one may sum up over many pure state trajectories finding the time evolution for an initially mixed state.
But with a temperature diffusive behavior already for pure states we are sure to find it also for an initially mixed state, even with reduced fluctuations.

\begin{figure}
  \centering 
  \includegraphics[width=8cm]{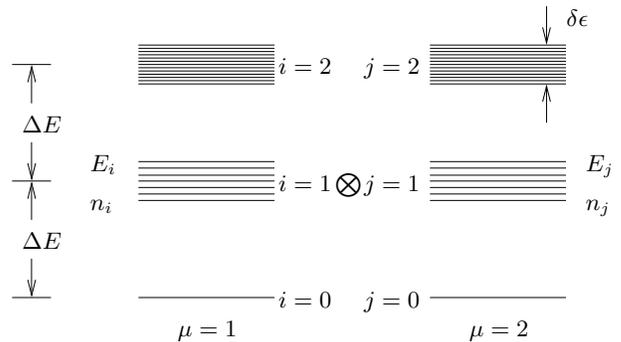}
  \caption{Heat conduction model with more exited bands: each band consists $n$ equally distributed levels ($N=2$ subunits).}
\label{fig:4}  
\end{figure}

Let us restrict ourselves in the following to only two subunits -- the proper minimal model as argued before. 
However, let us immediately generalize the model to a more complicated structure with more than one excitation band per subunit.
The results of HAM from the above considerations is valid also for this more general situation.
The model system is shown in \reffig{fig:4}.
The band index $i$ is reserved for the left system whereas $j$ belongs to the right system.
We take a product state $\ket{\psi}=\ket{\psi_1}\otimes\ket{\psi_2}$ as the respective initial state for all further investigations.
For the single subsystem states we choose a ``quasi'' thermal state with temperature $\beta_{\mu}$, i.e., a superposition of states according to the Boltzmann distribution. 
Here, we use the term ``quasi'' to label this special group of pure states, even if it is \emph{not} a mixed state as usual for thermal states. 
Nevertheless, the occupation probability of the energy eigen states refers to the Boltzmann distribution!
Thus, a mixture of many of such ``quasi'' thermal pure states may lead to the usual thermal state.
Furthermore, we treat the bands as ``quasi'' degenerate since the band width is much smaller than the gap $\delta\epsilon\ll\Delta E$.
Thus, the levels inside each band are equally occupied according to the temperature $\beta_{\mu}$.
The probability distribution of the first subsystem depending on the band energy $E_i$ and the appropriate number of states $n_i$ in the band is therefore given by
\begin{equation}
  \label{eq:37}
  P_{1}(E_i) = \frac{n_i\E^{-\beta_{1}E_i}}{Z(\beta_{1})}\;,
\end{equation}
with the partition function $Z(\beta_{1})$.
The second subsystem distribution is obtained simply by exchanging $i$ by $j$ and the respective quantities of the second subsystem.

The important quantity for all further considerations is the probability distribution of the whole system.
Since the total state is a product state the probability distribution of the complete system state $P_{ij}$ is simply the product of the single probabilities 
\begin{equation}
  \label{eq:38}
   P_{ij}
   = \frac{n_i\E^{-\beta_{1}E_i}}{Z(\beta_{1})}
     \frac{n_j\E^{-\beta_{2}E_j}}{Z(\beta_{2})}\;.
\end{equation}
However, it is convenient here to use a new set of coordinates for the combined spectrum in terms of the mean energy $E$ of a band and some energy displacement $\mathcal{E}$
\begin{equation}
  \label{eq:39}
  E           := E_i+E_j\;,\quad
  \mathcal{E} := \frac{1}{2}\big(E_i-E_j\big)\;.
\end{equation}
\begin{figure}
  \centering
  \includegraphics[width=8cm]{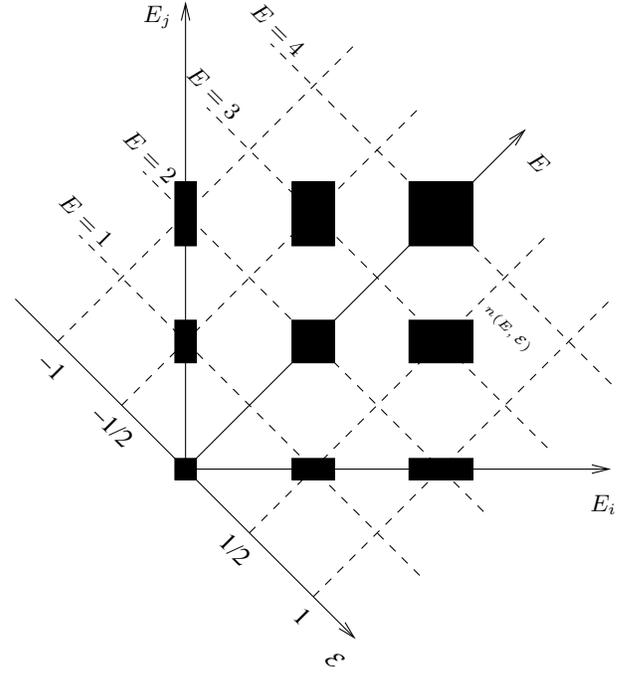}
  \caption{Alternative scheme to depict the complete multi-band spectrum: individual spectra of the two subunits on the $E_i$, $E_j$ axes respectively; transformation to the new parameters $(E,\mathcal{E})$; black bands indicate the complete spectrum. Examples: $(E,\mathcal{E})=(1,-1/2)$ are states, where the first system is in its first band, the second one in its ground state, $(E,\mathcal{E})=(2,0)$ are states, where both systems are in the first excited band etc.}
\label{fig:5}  
\end{figure}%
(All energies, also $E$ and $\mathcal{E}$, are measured in units of $\Delta E$.)
To support intuition we show both sets of coordinates in \reffig{fig:5}:
The full spectrum (black bands) can either be characterized in terms of the single subsystem spectra $(E_i,E_j)$ or by $(E,\mathcal{E})$.
The latter coordinates appear as a $45^{\circ}$-rotation of the original ones.
 
Introducing a small temperature difference $\Delta T$ around the mean temperature $T$ so that $\beta_{1/2}=(k_{\text{B}}(T\pm\Delta T/2))^{-1}$ and transforming to the new parameters (\ref{eq:39}) one finds the probability distribution
\begin{align}
  \label{eq:40}
  P(E,\mathcal{E})
  &= \frac{n(E,\mathcal{E})}{Z(\beta_{1})Z(\beta_{2})}
     \expfkt{-\frac{ET-\mathcal{E}\Delta T)}{k_{\text{B}}(T^2-(\Delta T/2)^2)}}
     \;,
\end{align}
with the number of states $n(E,\mathcal{E})$ in the band $(E,\mathcal{E})$.
Furthermore, in order to get a near equilibrium situation the temperature gradient $\Delta T$ has to be very small in comparison to the mean temperature $\Delta T\ll T$.
Therefore we expand the exponential function in (\ref{eq:40}) in terms of $\Delta T$ finding in first order
\begin{equation}
  \label{eq:41}
  P(E,\mathcal{E})
  \approx \frac{n(E,\mathcal{E})\E^{-\frac{E}{k_{\text{B}}T}}}{Z^2(T)}
    \Big(1+\frac{\mathcal{E}}{k_{\text{B}}T^2}\Delta T \Big)\;,
\end{equation}
where also the product of partition functions has been approximated in first order of the temperature gradient $\Delta T$ by $Z(\beta_{1})Z(\beta_{2})\approx Z(T)Z(T)=Z^2(T)$.

Again, the important quantity is the energy current between the two subsystems.
The total energy in the first subunit is now given by $U_{1} = \sum_i E_i P_i$, where $P_i$ is the time dependent probability distribution in the first system.
To get the respective first system probability distribution $P_i$ from the time dependent full system probability distribution $P_{ij}$ we have to trace over the second subsystem by summing over $j$, finding $U_{1} = \sum_{ij} E_i P_{ij}$.
According to this total energy in subsystem 1 the current along the chain reads
\begin{align}
  \label{eq:42}
  J &= \frac{1}{2}\sum_{ij}
       \left( E_i \dod{P_{ij}}{t} - E_j \dod{P_{ij}}{t} \right)\;,
\end{align}
or, switching to the new parameters (cf.\ (\ref{eq:39}))
\begin{equation}
  \label{eq:43}
  J = \sum_{E,\mathcal{E}} \mathcal{E} \;\dod{}{t} P(E,\mathcal{E})\;.
\end{equation}
For a single band model the sum over $E$ would break down and we recover the result (\ref{eq:35}).

The change of the probability $P(E,\mathcal{E})$ is again described by the rate equation according to the HAM.
Since the complete model decomposes into several energy subspaces, the derivation is essentially just as given in \refsec{sec:4}.
The respective energy transport coefficient results as a more complicated form depending on the band it belongs to ($\kappa=\kappa(E,\mathcal{E},\mathcal{E}')$). 
However, we require that $\kappa$ is symmetric under the exchange of $\mathcal{E}$ and $\mathcal{E}'$, since the energy transport from system 1 to system 2 should be as good as in the reverse direction.
Therefore we end up with the HAM rate equation 
\begin{equation}
  \label{eq:44}
  \dod{P(E,\mathcal{E})}{t} 
  = - \sum_{\mathcal{E}'} \kappa(E,\mathcal{E},\mathcal{E}')
      \Big(P(E,\mathcal{E})- P(E,\mathcal{E}')\Big)
\end{equation}
The fact that $\kappa$ depends on the respective band is already obvious from (\ref{eq:27}), which depends on the state density of the band.
Plugging in the rate equation (\ref{eq:44}) into the current (\ref{eq:43}) we find
\begin{equation}
  \label{eq:45}
  J = -\sum_{E,\mathcal{E},\mathcal{E}'} 
       \mathcal{E}\,\kappa(E,\mathcal{E},\mathcal{E}') 
       \Big(P(E,\mathcal{E})- P(E,\mathcal{E}')\Big)\;.
\end{equation}
Again, the current is simply a function of the actual state.

For a short time step we may use the first order approximation of the ``thermal'' initial state (\ref{eq:41}).
Since we are summing over $\mathcal{E}$ and $\mathcal{E}'$ the zeroth order terms according to (\ref{eq:41}) will cancel.
The remaining first order part
\begin{align}
  \label{eq:46}
   J =& -\sum_{E,\mathcal{E},\mathcal{E}'} 
       \frac{\kappa(E,\mathcal{E},\mathcal{E}')}{k_{\text{B}}T^2}
       \frac{\E^{-\frac{E}{k_{\text{B}}T}}}{Z^2(T)}
   \notag\\
       &\quad\Big(n(E,\mathcal{E})\mathcal{E}^2
       -n(E,\mathcal{E}')\mathcal{E}\mathcal{E}'\Big)
       \Delta T\;,
\end{align}
gives a heat current proportional to the temperature difference $\Delta T$.
By comparing this current with Fourier's Law (\ref{eq:2}), one can extract the heat conductivity coefficient as
\begin{equation}
  \label{eq:47}
  \kappa_{\text{th}}=
  \sum_{E,\mathcal{E},\mathcal{E}'} 
        \frac{\kappa(E,\mathcal{E},\mathcal{E}')\E^{-\frac{E}{k_{\text{B}}T}}}
             {k_{\text{B}}T^2Z^2(T)}
        \big(n(E,\mathcal{E})\mathcal{E}^2
        -n(E,\mathcal{E}')\mathcal{E}\mathcal{E}'\big)
\end{equation}
a quantity depending only on parameters of the respective model system and on temperature.
This is a very general formula for the heat conductivity derived from HAM, valid for the class of model systems depicted in \reffig{fig:4} with an arbitrary number of bands.

%
%
\section{Heat Transport Coefficient}
\label{sec:7}

Let us return in the following to the single band model as shown in \reffig{fig:1}, with the partition function of a single $n+1$-level subsystem $Z(T)=1+n \exp(-\Delta E/k_{\text{B}} T)$.
In this case HAM leads to the unique energy diffusion constant $\kappa(E,\mathcal{E},\mathcal{E}')=\kappa$ (see (\ref{eq:27})). 
Furthermore, in this single band model only the energy band $E=1\Delta E$ contributes to the conductivity $\kappa_{\text{th}}$, since for all other bands all possible $\mathcal{E}$ parameters are zero (cf. \reffig{fig:5}).
There are two possible values for the band $E=\Delta E$, namely $\mathcal{E}=\pm\Delta E/2$ and the number of levels in these two sub-bands are equal $n(\Delta E,\Delta E/2)=n(\Delta E,-\Delta E/2)=n$, due to the construction of the model.
One thus gets for the conductivity (\ref{eq:47})
\begin{equation}
  \label{eq:48} 
  \kappa_{\text{th}} =
  \frac{2\pi k_{\text{B}}\lambda^2n}{\hbar\delta\epsilon}
  \frac{n\E^{-\frac{\Delta E}{k_{\text{B}}T}}}
             {\left(1 + n \E^{-\frac{\Delta E}{k_{\text{B}}T}}\right)^2}
  \left(\frac{\Delta E}{k_{\text{B}}T}\right)^2\;.
\end{equation}

\begin{figure}
  \centering
  \includegraphics[width=8cm]{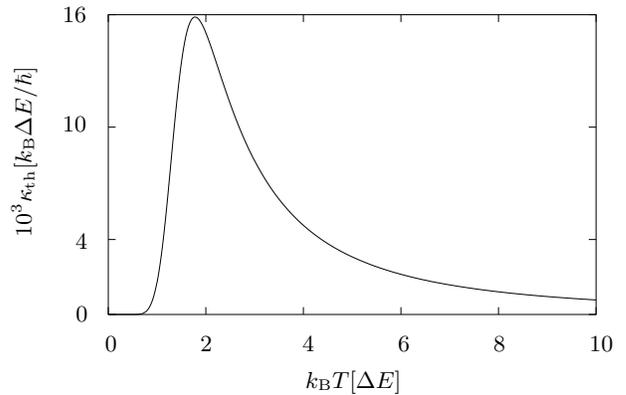}
  \caption{Heat conductivity (\ref{eq:48}) over temperature for a system with $n=500$, $\delta\epsilon = 0.05$ and $\lambda=5\cdot10^{-5}$.}
  \label{fig:6}
\end{figure}

The dependency of the transport coefficient $\kappa_{\text{th}}$ on temperature is displayed in \reffig{fig:6}.
It roughly looks as expected for typical isolating solids.
There is, however, no $\propto  T^3$ behavior for small temperatures, but note that this is only to be expected for three dimensional crystal lattices with a linear phonon dispersion relation for small momenta. 
The exponential behavior for small temperatures $\propto  \expfkttxt{-1/T}$ results from the gap in the system and is difficult to treat for small temperatures since it is an essential singularity (typical for systems with an energy gap). 
Furthermore we find a  $\propto T^{-2}$ behavior rather than the usual $\propto T^{-1}$ behavior in the high temperature limit. 
Again, the usual  $\propto T^{-1}$ is no rigorous general result, Peierls only states that the exponent should be between minus one and minus two.

By introducing the heat capacity, the above result for the heat conduction coefficient can further be simplified.
Starting with the canonical partition function of a single subunit $Z=1+n\E^{-\beta\Delta E}$ and the definition of the heat capacity in terms of the partition function 
\begin{equation}
  \label{eq:49}
  c = \frac{1}{k_{\text{B}}T^2}\pon{\ln Z}{\beta}{2}
\end{equation}
one finds for the heat capacity of one subunit
\begin{equation}
  \label{eq:50}
  c = k_{\text{B}}\frac{n\E^{-\frac{\Delta E}{k_{\text{B}}T}}}
           {\left(1 + n \E^{-\frac{\Delta E}{k_{\text{B}}T}}\right)^2}
  \left(\frac{\Delta E}{k_{\text{B}}T}\right)^2\;.
\end{equation}
From a comparison with (\ref{eq:48}) one derives
\begin{equation}
  \label{eq:51}
  \kappa_{\text{th}} 
  = \frac{2\pi\lambda^2n}{\hbar\delta\epsilon} c
  = \kappa c\;,
\end{equation}
i.e., the heat conductivity is given by the energy diffusion constant $\kappa$ multiplied by the heat capacity $c$ of a single subunit.

%
%
\section{Conclusion}
\label{sec:8}

Rather than considering heat conduction for a system coupled to two heat baths with different temperature in a stationary local equilibrium state, we have studied the time evolution of a closed one-dimensional chain of weakly coupled many level systems.
By an application of the new Hilbert space average method (HAM), we have been able to directly address a coarse-grained description. 
On this level of description the effective dynamics can be shown to be diffusive allowing us to  derive a theoretical expression for the \emph{energy diffusion coefficient} $\kappa$ from first principles (Schr\"odinger equation).
Furthermore, two conditions have been established which could help to classify whether a given model should show a statistical diffusive behavior or not.

The approximation involved (mainly the replacement of exact expectation values by a Hilbert space average) has been tested by a numerical comparison of the derived rate equation and the exact Schr\"odinger equation.
Surprisingly, we have found for the present model system that energy diffusion occurs right from the start and already far from equilibrium.

For an initial quasi thermal state featuring a small but finite temperature gradient we have been able to find a normal heat conducting behavior in terms of a finite \emph{heat conductivity} $\kappa_{\text{th}}$.
Even for a more complicated model system (several bands per subunit) it was possible to derive a theoretical expression for the transport coefficient in terms of the parameters of the Hamiltonian.
Finally, a reformulation of the transport coefficient has led to a connection with the energy transport coefficient and the heat capacity. 

What impact do those results for our ``design model'' have on real physical systems? 
For an application of HAM one has to organize the system as a ``grain-structure'' of weakly coupled subunits.
Then, for the emergence of normal transport, the local bands and their couplings that mediate the transport should obey the criterion (\ref{eq:29}) individually.
If this can be established HAM predicts that energy will diffuse from subunit to subunit. 

It has long since been appreciated that coarse-graining (i.e., tracing out irrelevant degrees of freedom) may change the effective dynamical laws even qualitatively.
An obvious example is encountered when some accelerated electron in a solid transfers energy and momentum to vibrational degrees of freedom thus giving rise to particle transport.
For the origin of heat transport in insulators a correspondingly simple picture seems to be missing though.

Here we have introduced a specific model class for which such a concept of information reduction can easily be incorporated.
We thus have been able to clearly specify the conditions under which the resulting effective dynamics can be characterized as normal energy diffusion.
Irreversibility and thermal resistance are thus traced back to a ``reduced description'': ``In reality'' the full dynamics is still controlled by the Schr\"odinger equation, as has explicitly been checked here for still small enough total quantum systems.

The desired structure may be achieved by coarse-graining periodic systems like spin chains, crystals, etc.\ into entities containing many elementary cells and taking the mesoscopic entities as the subunits proper and the effective interactions between the entities as the couplings. 
Increasing the ``grain size'' will result in higher state densities within the subunits and relatively weaker couplings such that above a certain minimum grain size the criteria (\ref{eq:29}) will eventually be fulfilled.
In this way a qualitatively new type of dynamical behavior can be seen to evolve as the observational level becomes more and more coarse-grained.

Of course, the resulting subunits cannot generally be expected to feature the same gapped spectral structure as our design model. 
But HAM should be applicable also to more realistic systems, and work in that direction is under way.

We thank H.-P.\ Breuer, M.\ Hartmann, M.\ Henrich, Ch.\ Kostoglou, H.\ Michel, H.\ Schmidt, M.\ Stollsteimer and F.\ Tonner for fruitful discussions. 
Financial support by the Deutsche Forschungsgesellschaft is gratefully acknowledged.


\end{document}